\documentclass[iop]{emulateapj}
\usepackage{amsmath}
\usepackage{amssymb}
\usepackage{epsfig}
\usepackage{subfigure}
\usepackage{natbib}
\usepackage{graphicx}
\usepackage{color}
\usepackage{longtable}
\usepackage{verbatim}

\shorttitle{Multi-level light curve modeling}
\shortauthors{Sanders et al.}

\newcommand{\tplatINDdistmean}{92} 
\newcommand{\tplatINDdiststd}{14} 
\newcommand{\tplatHIERdistmean}{90} 
\newcommand{\tplatHIERdiststd}{6} 
\newcommand{\Nsampcold}{12651} 
\newcommand{\Nsampcoldrate}{5.49} 
\newcommand{\Nsampcoldperchain}{395} 
\newcommand{\totalparams}{9,176} 
 
\newcommand{\NIIPtotalF}{76}

\newcommand{\PSOzerosixoneoneninesix}{PS1-10zu}

\newcommand{\PSOonetwozerotwoonefive}{PS1-11ai}

\newcommand{\PSOthreethreezerozerotwoseven}{PS1-11azd}

\newcommand{\PSOthreesevenzerofiveonenine}{PS1-12wn}

\newcommand{\PSOfourtwozerothreeninethree}{PS1-12bku}

\newcommand{\betatwotracepick}{PS1-12cey} 
\newcommand{\Nsampwarm}{11265} 
\newcommand{\rhPtracerhat}{1.77} 
\newcommand{\betatwotracepickrhat}{1.14} 
\newcommand{\NphotTot}{18,837} 
\newcommand{\NphotTotDet}{5,056} 
\newcommand{\tplatHIERdistmax}{112} 
\newcommand{\tplatAboveMaxPten}{60} 
\newcommand{\tplatAboveMaxPtwen}{30}

\newcommand{\tplatHIERdiststdmed}{33}

\newcommand{\gps}{\ensuremath{g_{\rm P1}}}
\newcommand{\rps}{\ensuremath{r_{\rm P1}}}
\newcommand{\ips}{\ensuremath{i_{\rm P1}}}
\newcommand{\zps}{\ensuremath{z_{\rm P1}}}
\newcommand{\yps}{\ensuremath{y_{\rm P1}}}

\newcommand{\PS}{\protect \hbox {Pan-STARRS1}}

\def\asec{\char'175 }

\begin{document}

\newcommand{\aCfA}{1}
\newcommand{\aUWar}{2}

\title{Unsupervised Transient Light Curve Analysis Via Hierarchical Bayesian Inference}
\author{
N.~E.~Sanders\altaffilmark{\aCfA}, 
M.~Betancourt\altaffilmark{\aUWar}, 
A.~M.~Soderberg\altaffilmark{\aCfA}
} 

\altaffiltext{\aCfA}{Harvard-Smithsonian Center for Astrophysics, 60 Garden Street, Cambridge, MA 02138 USA}
\altaffiltext{\aUWar}{Department of Statistics, University of Warwick, Coventry, UK}

\email{nsanders@cfa.harvard.edu}

\begin{abstract}
Historically, light curve studies of supernovae (SNe) and other transient classes have focused on individual objects with copious and high signal-to-noise observations.  In the nascent era of wide field transient searches, objects with detailed observations are decreasing as a fraction of the overall known SN population, and this strategy sacrifices the majority of the information contained in the data about the underlying population of transients.
A population level modeling approach, simultaneously fitting all available observations of objects in a transient sub-class of interest, fully mines the data to infer the properties of the population and avoids certain systematic biases.  We present a novel hierarchical Bayesian statistical model for population level modeling of transient light curves, and discuss its implementation using an efficient Hamiltonian Monte Carlo technique.
As a test case, we apply this model to the Type~IIP SN sample from the \PS\ Medium Deep Survey, consisting of \NphotTot\ photometric observations of \NIIPtotalF\ SNe, corresponding to a  joint posterior distribution with \totalparams~parameters under our model.  Our hierarchical model fits provide improved constraints on light curve parameters relevant to the physical properties of their progenitor stars relative to modeling individual light curves alone.  Moreover, we directly evaluate the probability for occurrence rates of unseen light curve characteristics from the model hyperparameters, addressing observational biases in survey methodology.
We view this modeling framework as an unsupervised machine learning technique with the ability to maximize scientific returns from data to be collected by future wide field transient searches like LSST.
\smallskip
\end{abstract}


\section{INTRODUCTION}
\label{sec:intro}

The majority of luminous transients in the universe are core-collapse supernovae (CC-SNe), marking the explosive deaths of massive stars \citep{Heger03,Smartt09}.  Stellar evolution theory, as well as both detailed observations of the explosive transient and fortuitous pre-explosion observations of the progenitor star, point to progenitor initial mass as the primary factor determining stars' eventual death state.  Metallicity, rotation rate, binarity, and other properties play important secondary roles, and permutations of these parameters are likely responsible for the extreme diversity of core-collapse supernovae phenomenology observed in the universe \citep{Heger03,Smartt09,Smith11,Ekstrom12,Jerkstrand13}.  The progenitor star mass distribution for each SN~type, as well as the distribution of these secondary factors, have far reaching implications throughout astrophysics, influencing the theory of stellar evolution \citep{Groh13}, galactic chemical evolution \citep{Nomoto13}, hydrodynamic feedback in galaxy formation \citep{Stilp13}, and astrobiology \citep{Lineweaver04}.

Studies of individual transients typically focus on well observed cases within each object class, capitalizing on the availability of detailed and high signal-to-noise observations to facilitate comparison to finely tuned hydrodynamic explosion simulations and analytic light curve models (e.g. \citealt{Mazzali03,Utrobin08}).  Syntheses of these observations, studies of large samples of SNe of a given class, are then typically composed of samples culled from these well observed cases (see e.g. \citealt{Nomoto06,Bersten09,Jerkstrand13}).  However, the properties of luminous and/or high signal-to-noise objects within a survey sample may be systematically different from their lower luminosity / signal-to-noise counterparts, and traditional targeted transient searches themselves are inherently biased towards particular SN progenitor properties like high metallicity \citep{SandersIbc,nes2010ay}.  To derive truly robust and unbiased inferences about SN progenitor populations, it is therefore necessary to study transient samples in a fashion as complete and observationally agnostic as possible.

Here we discuss a methodological framework for the simultaneous modeling of multi-band, multi-object photometric observations from wide field transient surveys, which addresses certain biasing factors inherent to transient searches.  This method is rooted in ``hierarchical'' and ``multi-level'' Bayesian analysis, where information about similar events within a sample is partially pooled through a hierarchical structure applied to the joint prior distribution (see \citealt{BDA3} and references therein; see \citealt{Mandel09} for applications to SN~Ia light curves).  We adopt Hamiltonian Monte Carlo as a computational technique to efficiently explore the high-dimensional and strongly correlated posterior distribution of this hierarchical model \citep{Betancourt13}.  The result of this modeling is simultaneous inference on physically-relevant light curve parameters describing individual objects in the sample, as well as the parameter distribution among the population, regularized by the application of minimal (``weakly informative'') prior information.

In Section~\ref{sec:model} we discuss the design and implementation of a hierarchical Bayesian model capable of simultaneously fitting large quantities of raw photometric data from wide field transient surveys to infer the population properties of the underlying SN sample.  We test this model with a sample dataset of Type~IIP SNe from the \PS\ (PS1) survey (Section~\ref{sec:data}), previously published in \cite{Sanders14IIP}.  We explore the results of this test in Section~\ref{sec:results}, including comparison with inferences drawn from traditional modeling based on fits to individual light curves.  We discuss the implications of this methodology for future wide field transient surveys in Section~\ref{sec:disc} and conclude in Section~\ref{sec:conc}.

\section{MODEL DESIGN}
\label{sec:model}

We have designed a hierarchical Bayesian generalized linear model (GLM) to simultaneously describe the individual multi-band light curves of a set of optical transients, and the population distribution of their light curve parameters.  Due to the nature of the sample dataset we discuss in Section~\ref{sec:data}, the non-linear link function of our GLM is tailored for Type~IIP SNe (see Section~\ref{sec:model:lc}), but the hierarchical structure of the model is generalizable to any transient class.

Type~IIP SNe are particularly apt for a hierarchical modeling approach because their long lived light curves reduce the likelihood of any individual object to have fully identifiable light curve parameters.  In particular, because the plateau phase of the SN~IIP light curve has a duration ($\sim3$~months) similar to the length of observing seasons for typical pointings of ground based telescopes, individual light curves are typically incomplete.  The detected SNe~IIP have often exploded between observing seasons, when their field is behind the sun, or their field sets before the plateau phase has ended.  As a result, individual objects in the data set do not have the temporal coverage needed to fully identify their light curve parameters.  Partial pooling among objects in the sample can compensate, helping to identify unconstrained parameter values for individual objects, while applying information from well-constrained parameters of the individual light curves to all other objects in the sample.

\subsection{Light curve model}
\label{sec:model:lc} 

\begin{figure}
\plotone{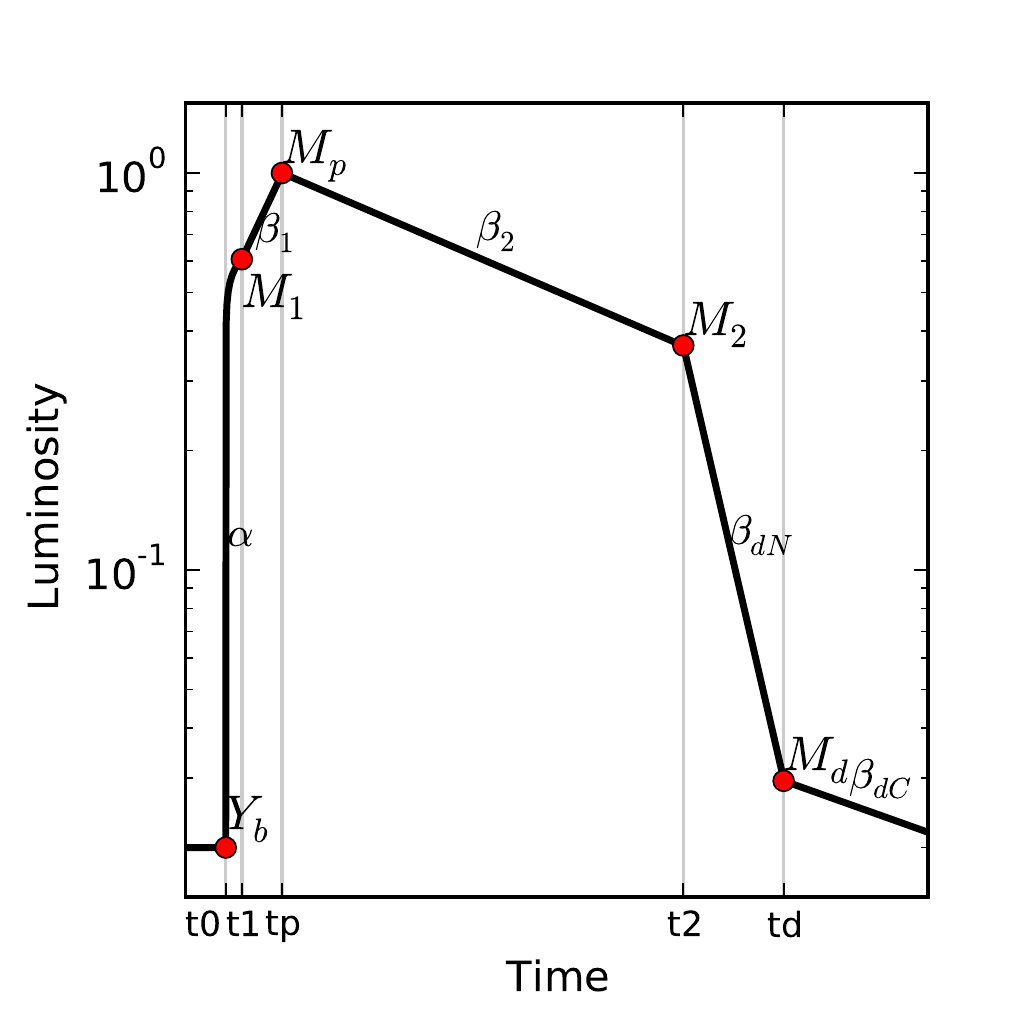}
\caption{\label{fig:modschem}Schematic illustration of the 5-component SN~II light curve model defined in \cite{Sanders14IIP}.  The gray vertical lines denote the epochs of transition ($t_x$) between the piecewise components of the model.  The background level ($Y_b$) and turnover fluxes ($M_x$) are marked and labeled (red points).  The power law ($\alpha$) and exponential ($\beta_x$) rate constant for each phase is labeled adjacent to the light curve component.}
\end{figure}

We have designed a physically motivated parameterized model for the SN~IIP light curve, composed of 5 piecewise power law and exponential segments.  The model is fully specified by a set of 12 independent parameters per optical filter and an explosion date ($t_0$).  These parameters are 4 time durations ($t_1,t_p,t_2,t_d$) defining the knot locations of the segments, 5 rate parameters describing the slope of each light curve segment ($\alpha,\beta_1,\beta_2,\beta_{dN},\beta_{dC}$), a luminosity scale ($M_p$), a background level ($Y_b$) for the photometric data, and an intrinsic scatter ($V$) to encompass deviation from the model.  The light curve model and its primary parameters are illustrated in Figure~\ref{fig:modschem}; a full mathematical description of the light curve model is given in \cite{Sanders14IIP}.

\subsection{Hierarchical structure}
\label{sec:model:ml} 

\begin{figure*}
\plotone{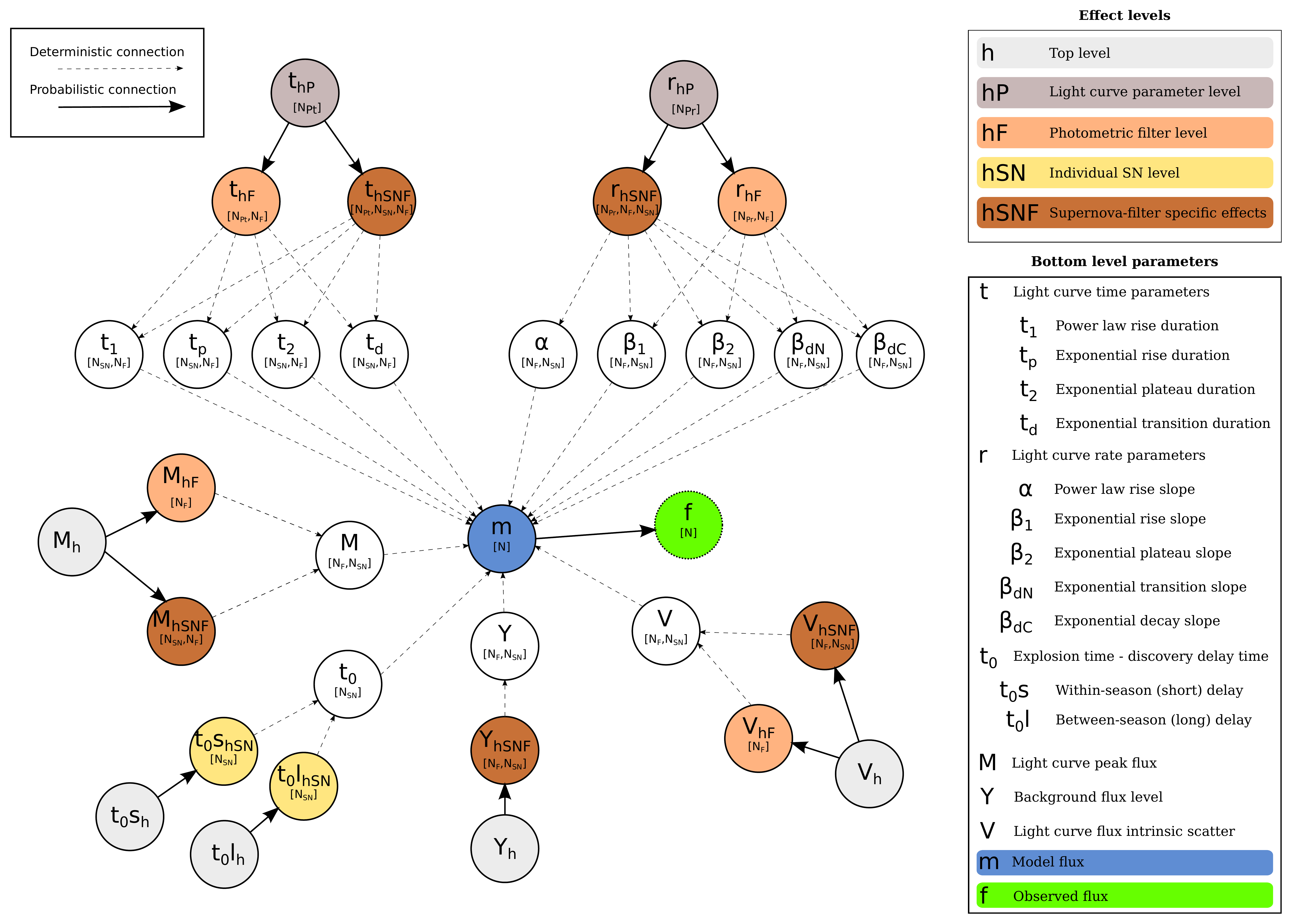}
\caption{\label{fig:dagm7}Directed Acyclic Graphical representation of the hierarchical structure of the multi-level light curve model parameters.  The color coding of the effects levels and the meanings of the bottom level parameters are given in the key at right.  The bracketed numbers indicate the dimensionality (in terms of the number of objects, $N_{SN}$, the number of filters, $N_F$, and the number of parameters in the time and rate groups, $N_{Pt}$ and $N_{Pr}$) of each parameter matrix.}
\end{figure*}

We allow for partial pooling between the light curve parameters of this 5-component model using a linear hierarchical structure.  This structure is illustrated in Figure~\ref{fig:dagm7}.  For the time, rate, luminosity scale, and intrinsic scatter parameters of the model, this structure includes levels for individual photometric filters, individual SNe-filter combinations, and a top-level parameter; a 3-level structure.  For the explosion date ($t_0$) parameter, which is not defined per-filter, we use a modified two-level structure.  The structure includes separate hyperprior distributions for objects discovered within and between observing seasons, which will have significantly different delay times between explosion and the epoch of detection.  We do not adopt a hierarchical structure for the background level $Y_b$ parameter, which should nominally be $0$ except in the presence of artifacts among the PS1 template images used in difference imaging.

In effect, this structure means that a single top level value is drawn for each rate parameter; separate filter-level rate parameters are drawn for each of the $grizy$ filters from the hyperprior distribution specified by the top level value; and bottom level rate parameters for each SN-filter combination are drawn from the hyperprior specified by the filter-level draw.  In practice, this ``centered'' multi-level parameterization is non-optimal, because it introduces significant correlations between the hyperparameters in the model that decrease the efficiency of the MCMC sampler.  Instead we use a modified, ``non-centered parameterization,'' where correlations between hyperparameters are exchanged for correlations between hyperparameters and data.  This is a general technique applicable for any distribution in the location-scale family, and optimal when the data poorly identify the parameter values \citep{Papaspiliopoulos07,Betancourt13}.  We therefore adopt normal hyperprior distributions for all the location hyperparameters in our model, and half-cauchy distributions for all scale parameters (including hyperprior width parameters; \citealt{Gelman08}).

The hierarchical modeling framework largely eschews the specification of prior information, instead allowing the model to set its own hyperprior distributions learned from the data during fitting.  We view this process, as applied to transient optical light curve studies, as a form of unsupervised machine learning.  In effect, the model is learning the shape and range of variation among light curves within the transient class, and applying that information to optimally interpret individual light curves.

However, it is necessary to set prior distributions for the top level hyperparameters, and we adopt weakly informative priors except where needed to enforce regularization of the light curve model.  In particular, we assign mean values for the normal prior distribution on the filter-level parameter ($t_{hF,t_p}$) controlling the plateau phase rise time ($t_p$) to specify the within filter variation observed in \cite{Sanders14IIP}.  We do the same for the filter-level priors controlling the plateau phase rise and decay rates ($\beta_1$ and $\beta_2$).  We specify the prior on the explosion date hyperparameters with means of 1 and 100~days for the within- and between-season objects, respectively.  We use a restrictive $\rm{cauchy}(0.001)$ hyperprior for the top-level intrinsic scatter parameter ($V_h$) to regularize its ability to dominate the likelihood evaluation.  We note that narrow hyperprior distributions are needed here because the hierarchical model exponentially amplifies variances.  Prior information is therefore needed to ensure a reasonable range of variation of the top level parameters and to avoid numerical overflow during sampling.  The model then fits optimal values for each of these hyperparameters given the likelihood for the data, and these priors serve largely to regularize the results.

\subsection{Stan implementation}
\label{sec:model:stan} 

To sample from this model posterior, we employ the C++ library \textit{Stan} \citep{STAN}, which implements the adaptive Hamiltonian Monte Carlo (HMC) \textit{No-U-Turn Sampler} (NUTS) of \cite{NUTS}.  HMC is advantageous for inference on high dimensional multi-level models, because it capitalizes on the gradient of the posterior to efficiently traverse the joint posterior despite the presence of the highly correlated parameters inherent to hierarchical models \citep{Betancourt13}.  In practice, HMC will achieve a significantly higher effective sample size ratio (i.e. lower autocorrelation in the trace) than traditional Gibbs samplers for models with highly correlated parameters \citep{Betancourt13,stan-manual:2014}.

NUTS operates in two phases; ``adaptation'' and ``sampling.''  During adaptation, the algorithm automatically tunes the temporal step size which controls the discretization of the Hamiltonian \citep{hgelman2013}.  Additionally, the algorithm estimates a diagonal HMC mass matrix during adaptation, which effectively scales the global step size to the optimal value for each parameter (we do not configure \textit{Stan} to estimate the full, ``dense'' mass matrix given the significant additional computational overhead).  During the sampling phase, the step size and mass matrix are fixed.

We use \textit{Stan} to construct {32} independent MCMC chains from the posterior distribution of the model.\footnote{The full \textit{Stan} code for our statistical model is discussed in Appendix~\ref{ap:stan}.}  We have used the Harvard Faculty of Arts and Sciences ``Odyssey'' Research Computing cluster to run these chains in parallel, running for the cluster's maximum job execution time of 3~days per chain, for a total utilization of $2,304$~cpu~hrs.  Given our total yield of \Nsampcold\ samples, this represents an average chain length of \Nsampcoldperchain~samples and an effective sampling rate of {\Nsampcoldrate}~samples per hour per chain.  For the purposes of convergence testing (Section~\ref{sec:res:converge}), we consider the full chains including adaptation phase.  For the purposes of light curve modeling, we exclude the adaptation phase as well as the first 20~iterations of the sampling phase, yielding \Nsampwarm\ total samples from the approximate posterior stationary distribution.

\begin{figure}
\plotone{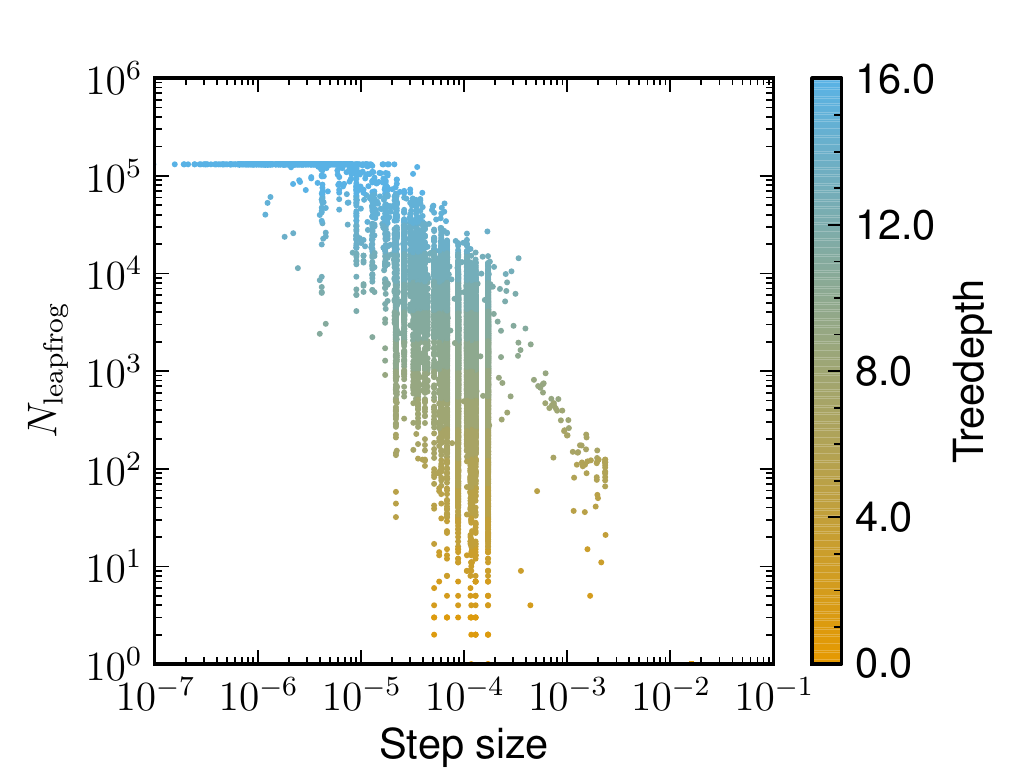}
\caption{\label{fig:treedepth}Illustration of the complexity of the HMC sampling procedure.  The figure compares the HMC Hamiltonian discretization step size to the number of leapfrog steps ($N_{\rm{leapfrog}}$) needed per iteration.  All samples from the combined MCMC chain, including adaptation steps, are shown.  The color coding shows the NUTS treedepth (key at right).  Vertically-correlated features indicate adapted chains (proceeding with fixed step size).}
\end{figure}

The high computational cost of sampling from the model posterior distribution is due to the small HMC step size emerging from the NUTS adaptation.  Figure~\ref{fig:treedepth} illustrates this effect, comparing the Hamiltonian discretization step size to the number of leapfrog steps per iteration as the step size varies during NUTS adaptation.  As the step size decreases, the number of leapfrog steps needed (the number of posterior calculations, and therefore the execution time) grows exponentially.  The horizontal feature at the top of the this figure illustrates saturation of the leapfrog algorithm tree depth, suggesting that yet smaller step sizes may be needed to optimally sample from the posterior.  However, given the onerous computation time required to iterate the NUTS algorithm (which is not immediately parallelizable) at the selected maximum tree depth ($\gtrsim1$~hour of CPU time at the maximum tree depth of 16), we have elected not to increase the maximum tree depth. As a result, the HMC sampler could potentially become stuck in local minima of the multi-dimensional posterior, biasing the resulting samples away from the tails of the true joint posterior distribution.

\section{SAMPLE DATA}
\label{sec:data}

\subsection{\PS\ Optical Observations}
\label{sec:data:PS1} 

Our Type~IIP supernova light curve sample is selected from four years of systematic Medium Deep Field observations by the \PS\ (PS1) telescope, as described in \cite{Sanders14IIP}.  PS1 is a high-etendue wide-field imaging system, designed for dedicated survey observations and located on a mountaintop site in the Hawaiian island chain.  Observations are conducted remotely, from the University of Hawaii--Institute for Astronomy Advanced Technology Research Center (ATRC) in Pukalani.  A complete description of the PS1 system, both hardware and software, is provided by \cite{PS1}. The 1.8~m diameter primary mirror, $3.3^\circ$ field of view, and other PS1 optical design elements are described in \cite{PS1opt}; the array of 0.258\asec pixel detectors, and other attributes of the PS1 imager is described in \cite{PS1cam}; and the survey design and execution strategy are described in \cite{PS_MDRM}.  The PS1 Medium Deep Survey (MDS) consists of 10 pencil beam fields observed with a typical cadence of 3~d in each filter.

The PS1 observations are obtained through a set of five broadband filters, which we refer to interchangeably as as \gps, \rps, \ips, \zps, and \yps or simply $grizy$ \citep{PS1cal}. MDS achieves a $5\sigma$ depth of $\sim23.3$~mag in $griz$ filters, and $\sim21.7$~mag in the $y$-filter (with observations taken near full moon).  Photometry presented here is in the ``natural'' PS1 system, $m = −2.5 \log(\rm flux) + m^\prime$, with a single zero-point adjustment $m^\prime$ made in each band to conform to the AB magnitude scale \citep{Schlafly12,Tonry12,Magnier13}.\footnote{The magnitudes quoted throughout this paper are in the AB system, except where explicitly noted.}  We assume a systematic uncertainty of 1\% for our PS1 observations due to the asymmetric PS1 point spread function and uncertainty in the photometric zero-point calibration \citep{Tonry12}.  The standard reduction, astrometric solution, and stacking of the nightly images is done by the Pan-STARRS1 IPP system \citep{PS1_IPP,PS1_astrometry}, and the nightly MDS stacks are processed through a frame subtraction analysis using the \textit{photpipe} image differencing pipeline \citep{Rest05,Scolnic13}.

\label{sec:data:sample} 

We adopt the final spectroscopic SN~IIP sample from \cite{Sanders14IIP}, including all objects sub-classified using the Support Vector Machine machine learning classification method therein.  This sample consists of \NphotTot\ total photometric data points, including \NphotTotDet\ robust detections, for \NIIPtotalF\ SNe~IIP in the $grizy$ filters.  We note that the photometric observations which are not robust detections still play a significant role in the likelihood of our model, serving to constrain the rise time and decay rate parameters of the model, as well as directly identifying the background parameter $Y_b$.  The particular transients included in this sample and their properties are described in \cite{Sanders14IIP}.

\subsection{Posterior Probability Convergence}
\label{sec:data:logprob} 

The HMC algorithm quickly and efficiently converges on a maximal value of the global posterior probability for the model by identifying optimal values for each bottom level light curve parameter for all the SNe and for the hyperparmeters.  Given that the global model for the PS1~SNe~IIP sample has a total of \totalparams\ individual parameters, this fast convergence is a significant testament to the efficiency of HMC as an optimization engine for high-dimensional functions.

\begin{figure}
\plotone{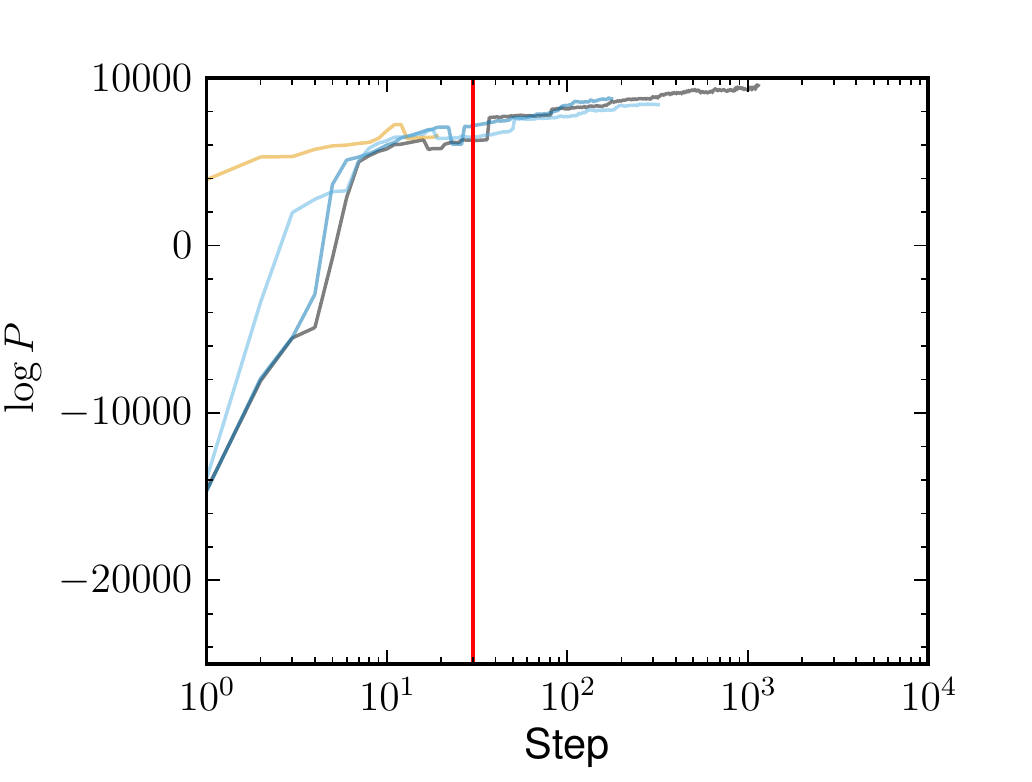}
\caption{\label{fig:trace:logprob}Markov Chain Monte Carlo trace of the global posterior probability from the hierarchical model fit, showing the fast convergence of the HMC algorithm to near the optimal parameter values.  A random subset of chains from the hierarchical model fit are shown (different lines).  The red vertical line marks the end of the NUTS adaptation phase, at which point the HMC step size is fixed.  The probability shown on the $y$-axis is not normalized and therefore has arbitrary units.}
\end{figure}

Figure~\ref{fig:trace:logprob} shows the posterior probability evolution of the Markov chains as the NUTS sampler adapts and then reaches the sampling phase.  Chains typically converge near the maximum achievable posterior probability during our warmup period of only 30 iterations.

\section{RESULTS}
\label{sec:results}

\subsection{Sampling Characteristics and Fit Convergence}
\label{sec:res:converge} 

\begin{figure}
\plotone{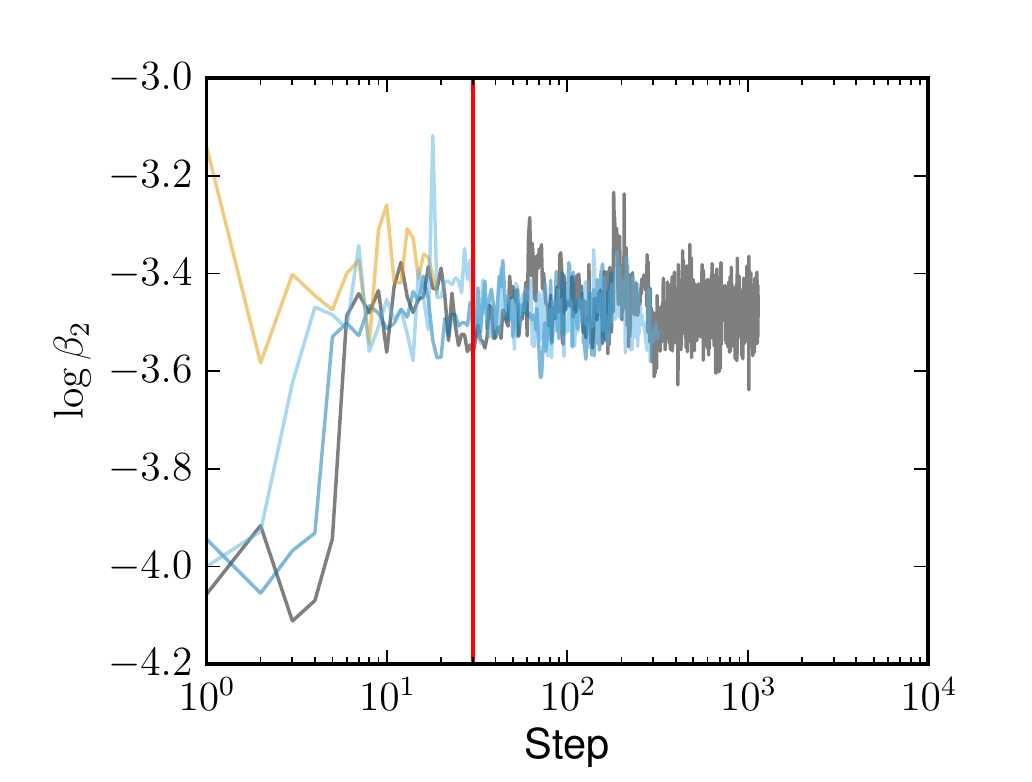}
\caption{\label{fig:trace:beta2}Illustration of convergent HMC sampling for the well identified bottom-level model parameters.  Like Figure~\ref{fig:trace:logprob}, but for the bottom level parameter $\beta_2$ for a randomly selected SN ($r$ band).}
\end{figure}

Figure~\ref{fig:trace:beta2} shows the MCMC trace for a well identified bottom level parameter; the values drawn from the HMC algorithm for the plateau phase decline rate ($\beta_2$) of an object (\betatwotracepick) with sufficient $r$-band photometry to constrain this phase of the light curve.  The sampler moves quickly in this dimension, with low autocorrelation between samples, and the parameter is acceptably convergent (with potential scale reduction factor $\hat{R}=\betatwotracepickrhat$).

\begin{figure}
\plotone{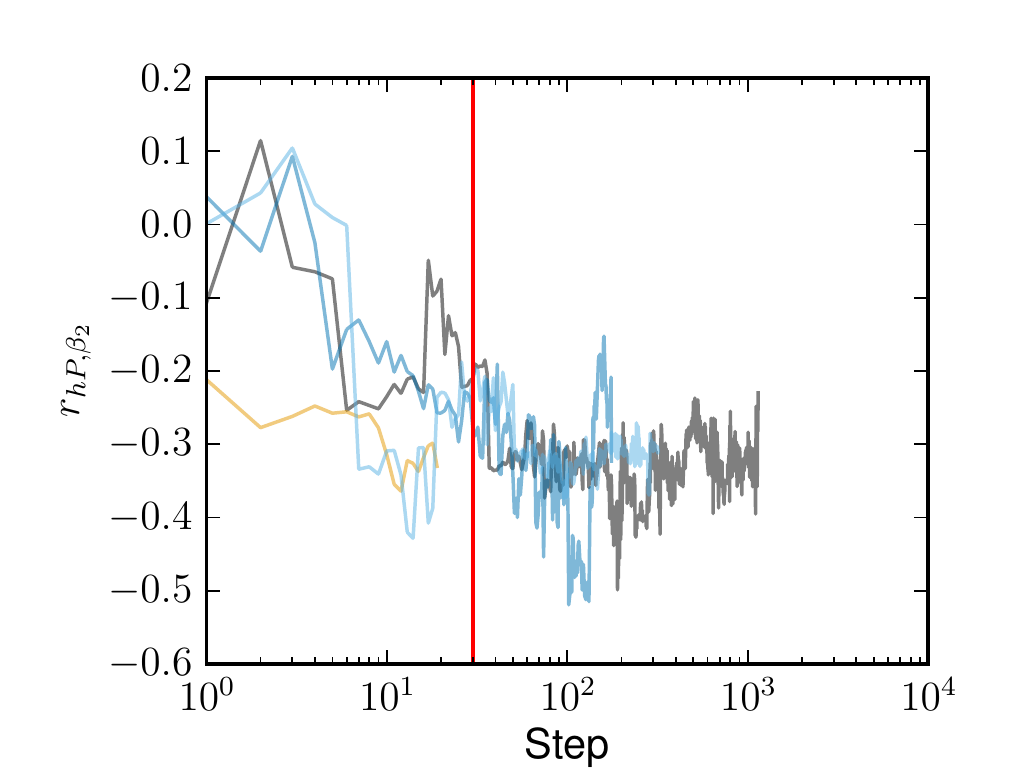}
\caption{\label{fig:trace:rhP}Illustration of slow moving HMC sampling for the higher level model parameters.  Like Figure~\ref{fig:trace:logprob}, but for the hyperparameter $r_{hP,\beta_2}$, the top level parameter controlling the $\beta_2$ decay rates.}
\end{figure}

In contrast, Figure~\ref{fig:trace:rhP} shows the trace for a moderately well identified top level parameter ($r_{hP}$), controlling the global plateau phase decay rate across all filters.  The trace indicates that the sampler is moving more slowly in this dimension, with significant autocorrelation between samples.  This top level parameter has apparently not yet converged ($\hat{R}=\rhPtracerhat$).  This suggests that additional sampling is needed to achieve a desirable level of convergence among some hyperparameters, but the computational cost is prohibitive at this time; we discuss alternative methods for achieving convergence with HMC in Section~\ref{sec:disc}.

\begin{figure}
\plotone{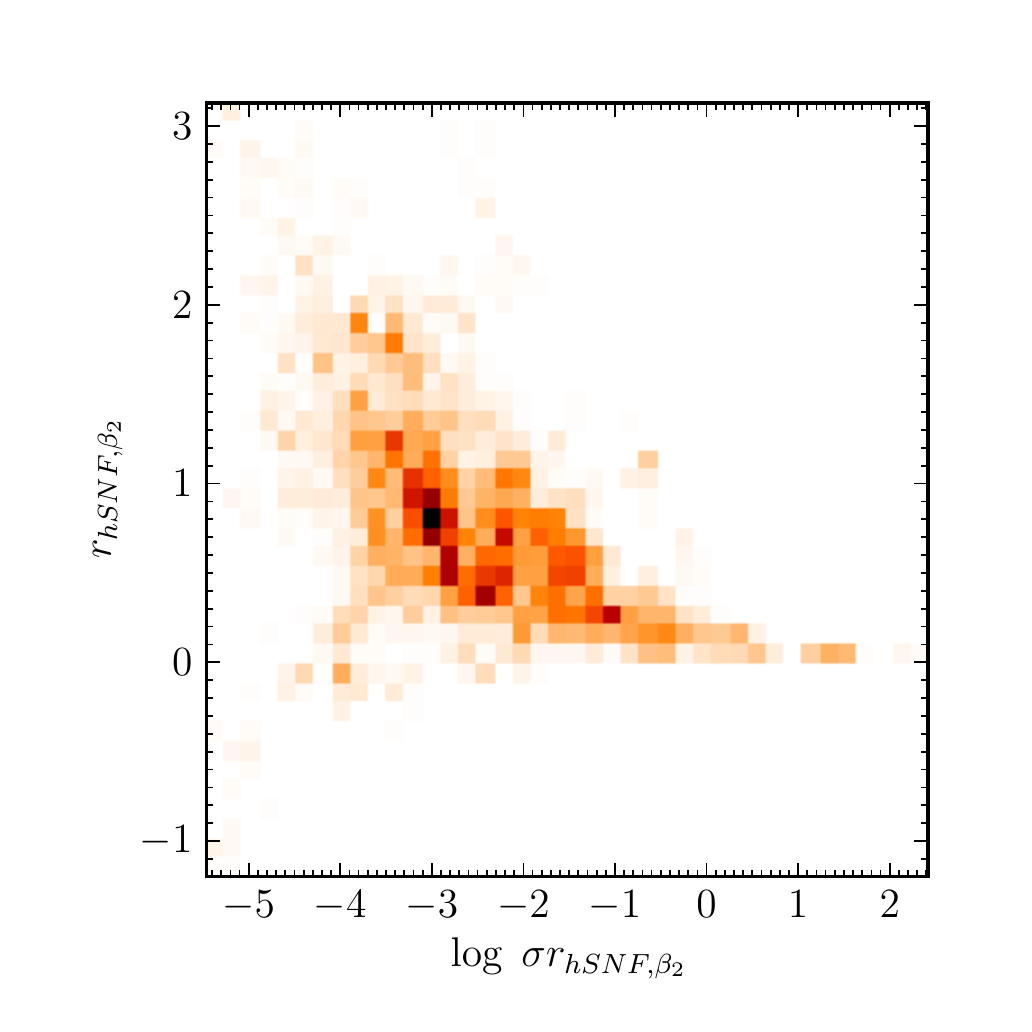}
\caption{\label{fig:joint:rhSNF}Illustration of the high curvature in the hierarchical model posterior.  A histogram of samples from a slice of the joint posterior distribution is shown, along the dimensions of the SN-filter level mean and width (on a log scale) hyperparameters for the plateau phase decay rate ($r_{hSNF,\beta_2}$).}
\end{figure}

\begin{figure}
\plotone{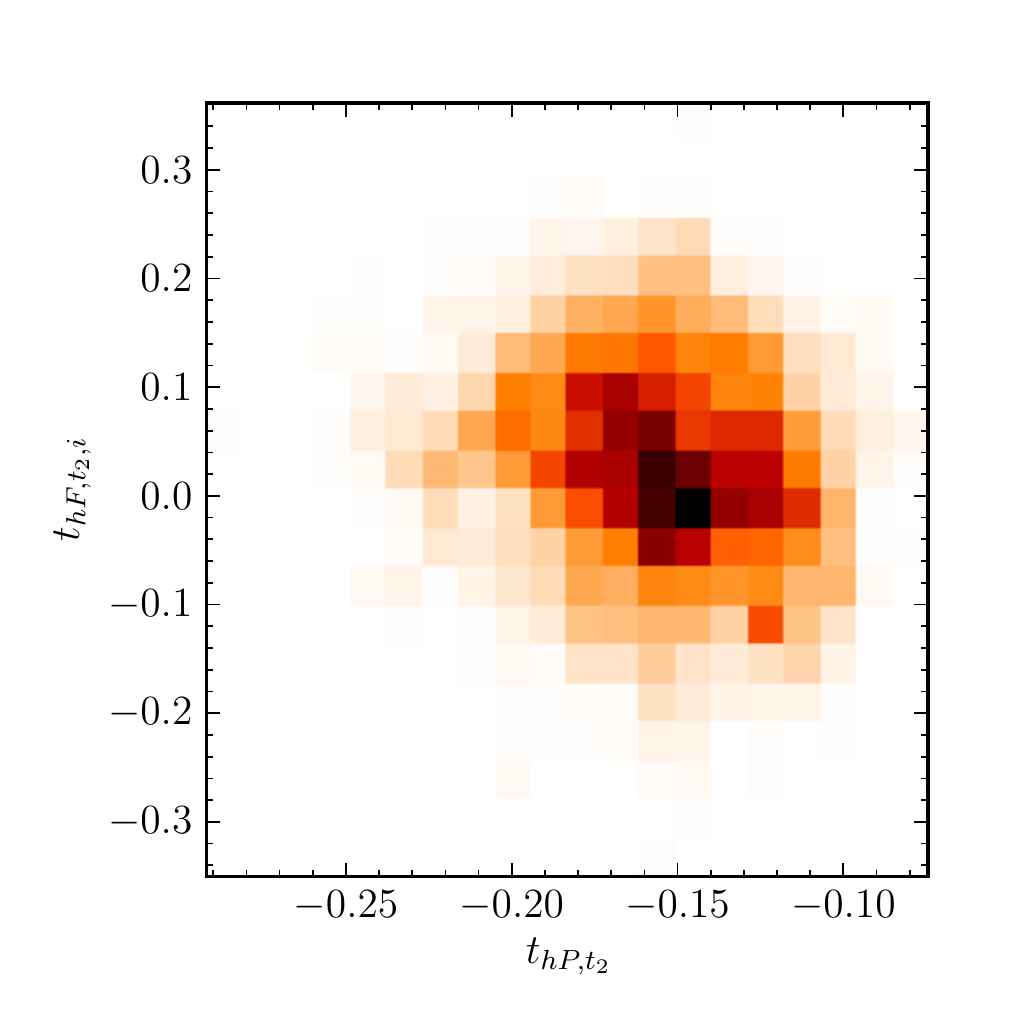}
\caption{\label{fig:joint:thPthF}Illustration of the effectiveness of the non-centered parameterization in removing dependency between hyperparameters.  Like Figure~\ref{fig:joint:rhSNF}, but showing a slice along the dimensions of the top level and filter-level hyperparameters for the plateau phase time duration ($t_{hP,t_2}$ and $t_{hP,t_2,i}$).}
\end{figure}

The origin of the convergence challenges facing the HMC algorithm are illustrated in Figure~\ref{fig:joint:rhSNF}.  The Figure, showing a slice from the joint posterior distribution, illustrates the high correlation between the hierarchically linked parameters in the model.  In contrast, Figure~\ref{fig:joint:thPthF} shows a slice of the joint posterior along the dimensions of the top level and filter-level hyperparameters for the plateau phase time duration.  Dependence between these hyperparameters was obviated via selection of the non-centered parameterization (Section~\ref{sec:model:ml}) and, indeed, their marginal posteriors have very low correlation.

\subsection{Posterior Predictive Check Comparison}
\label{sec:res:PPC} 

\begin{figure}
\plotone{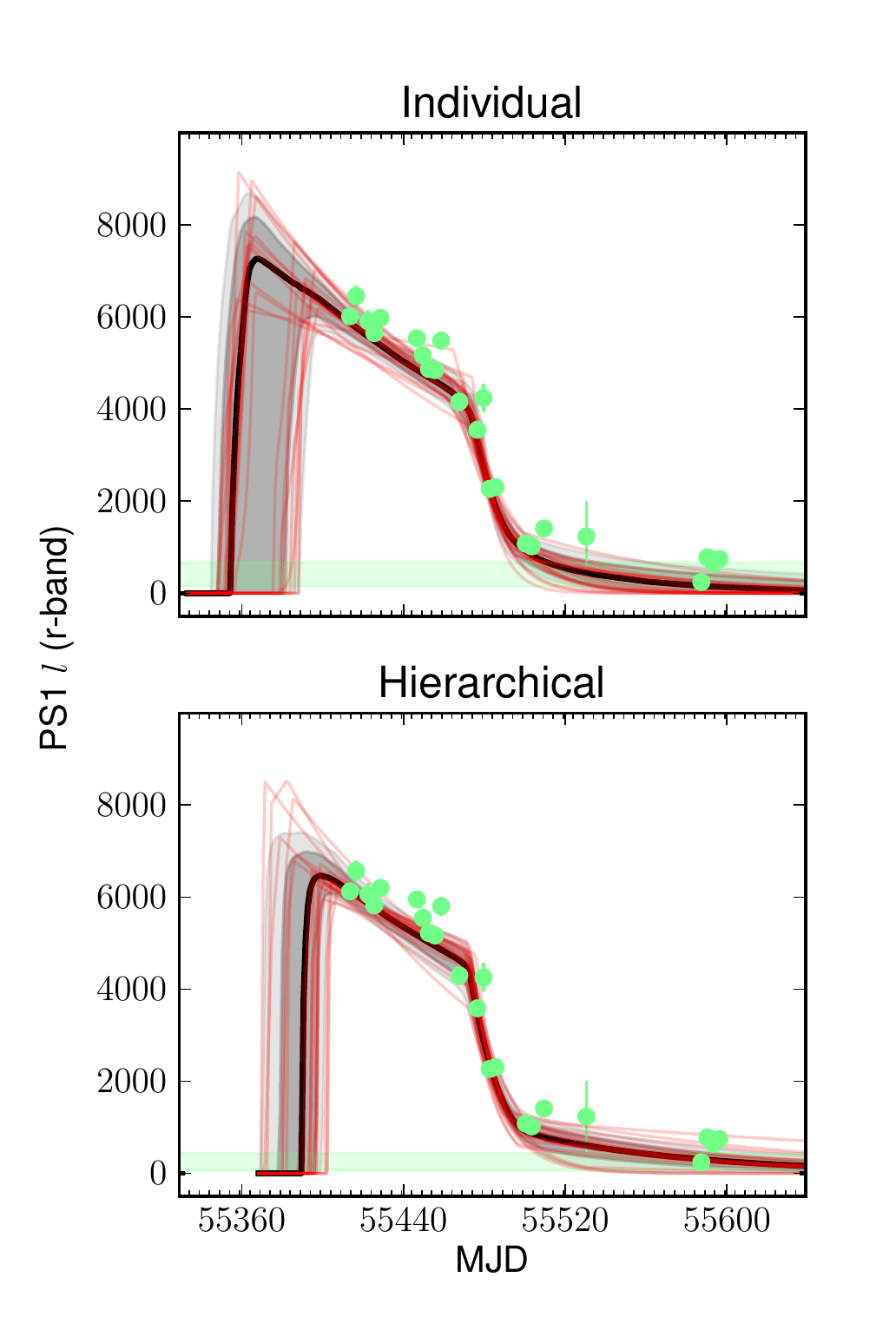}
\caption{\label{fig:PPC:lq}Comparison of posterior predictive checks between the individual light curve fit of \cite{Sanders14IIP}, above, and the hierarchical model fit, below, for the SN {\PSOzerosixoneoneninesix}.  Each plot shows the $r$-band photometry for the SN (green points and errorbars), which has detection only of the final phases of the light curve because the object exploded between observing seasons.  The units of luminosity, $l$, are described in \citep{Sanders14IIP}.  The shaded areas show the $1$ and $2~\sigma$ confidence intervals for the posterior predictive distribution of the 5-component light curve model (see Section~\ref{sec:res:PPC}), and the solid black line shows the median of that confidence interval.  The red lines show a random subset of sampled light curve models corresponding to parameter values from the MCMC chain.  The horizontal green bar shows the range fitted for the zero-point luminosity offset.}
\end{figure}

We validate the success of our model in describing the light curve behavior of objects in our SN~IIP sample using posterior predictive checks \citep{BDA3}, comparing the distribution of luminosities predicted under our fitted light curve model to the observed photometric data.  Figure~\ref{fig:PPC:lq} shows a posterior predictive check for {\PSOzerosixoneoneninesix}, whose poor temporal coverage illustrates the strengths of the hierarchical model.  The figure compares the $r$-band light curve fit for this object to the fit under the individual-level model presented in \cite{Sanders14IIP}, which uses an identical 5-component light curve model, but does not make use of no partial pooling between SNe.  The hierarchical fit achieves significantly greater constraints on the parameters describing the rising phases of the SN, resulting in a much tighter distribution of explosion dates and plateau durations (a parameter critical for physical inference on the progenitor star).  The improvement is due to the strongly identified plateau duration hyperparameters (Section~\ref{sec:res:pop}).  In the individual-level model, a weakly informative prior distribution was established for this parameter based on the theoretically predicted range of plateau duration variation; in the multi-level model, the hyperparameters are inferred from the data themselves, resulting in much stronger prior information at the individual SN level.  That the fit to the later phases of the light curve, where the data are strongly identifying, is indistinguishable from the fit obtained in the individual model is validation of the unbiased performance of the hierarchical model.

\begin{figure*}
\plotone{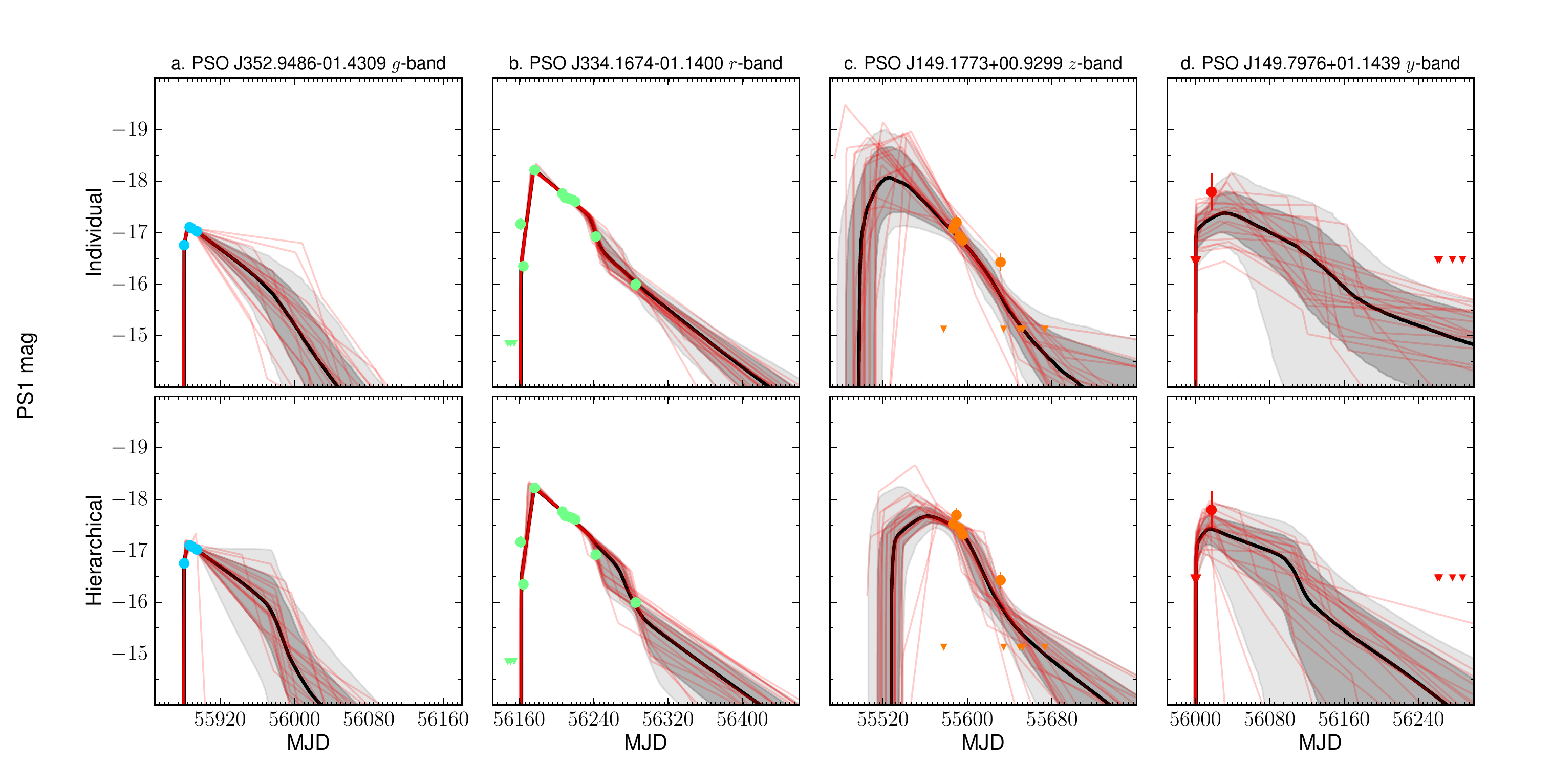}
\caption{\label{fig:PPC:hq}Illustration of the advantages provided by the hierarchical model for fitting under different data coverage scenarios, relative to individual light curve fitting.  Like Figure~\ref{fig:PPC:lq}, but on the absolute magnitude scale and for a variety of PS1~SNe and photometric filters.  The individual model fits are shown above, the with the corresponding hierarchical fits below.  Triangles represent the locations of photometry that are not robust detections; they are plotted at the $y$-axis position of the limiting PS1 magnitude of the filter, though the actual flux and uncertainty of these points is used in the Bayesian likelihood calculation.}
\end{figure*}

Figure~\ref{fig:PPC:hq} shows a comparison of fits for several additional objects, illustrating features of the hierarchical modeling framework under different data scenarios, which we describe here.  {\PSOthreethreezerozerotwoseven} was observed only during the rise and initial stages of the plateau phase.  The posterior predictive luminosity distribution of the individual and hierarchical models are similar, but the plateau decline phase duration ($t_2$) parameter is much more constrained in the hierarchical model (Figure~\ref{fig:PPC:hq}~a).  

{\PSOfourtwozerothreeninethree} (Figure~\ref{fig:PPC:hq}~b) was observed from explosion through the final, radioactive decay-dominated phase.  Generally, this case confirms that, where the data is strongly identifying, the hierarchical model produces fits in agreement with the individual-level model.  Interestingly, for this object there is a $r$-band photometric observation with relatively high uncertainty at $\sim+80$~days, which introduces a degeneracy in the posterior---whether this point should be assigned to the plateau or transition phase of the light curve.  The fit for this SN under the hierarchical model looks similar to the individual-level model fit, exploring both forks of the degeneracy.  However, the two fits favor opposite sides of the fork. The individual fit maximizes the likelihood of the $r$-band photometry for the object alone, placing the point on the transition phase, while the hierarchical fit prefers the solution where the point falls on the plateau.  Because the fork favored by the hierarchical model is more consistent with the modeled distribution of plateau durations among SNe~IIP (based on partial pooling from the other objects in the sample), it is the more well justified solution.

In Figure~\ref{fig:PPC:hq}~c, the hierarchical $z$-band fit for the SN {\PSOonetwozerotwoonefive} is much more highly regularized to match the shape of other $z$-band light curves than the individual fit.  This leads to significantly improved constraints on the peak magnitude and plateau duration for this object.  For {\PSOthreesevenzerofiveonenine} ($y$-band; Figure~\ref{fig:PPC:hq}~d), only 1 robust photometric detection is available, and all the photometry is highly uncertain ($\delta m\sim0.5$~mag at $1\sigma$).  The individual light curve fit in this case is very poorly regularized, obeying the peak magnitude suggested by the detection and the limits suggested by the non-detections, but otherwise has very poorly constrained light curve properties like plateau duration and decline rate.  The hierarchical fit in this case is far superior in regularization, showing a characteristic $y$-band SN~IIP light curve shape matched to the available photometry.

\subsection{Population Parameter Distribution Characteristics}
\label{sec:res:pop}

\begin{figure*}
\plotone{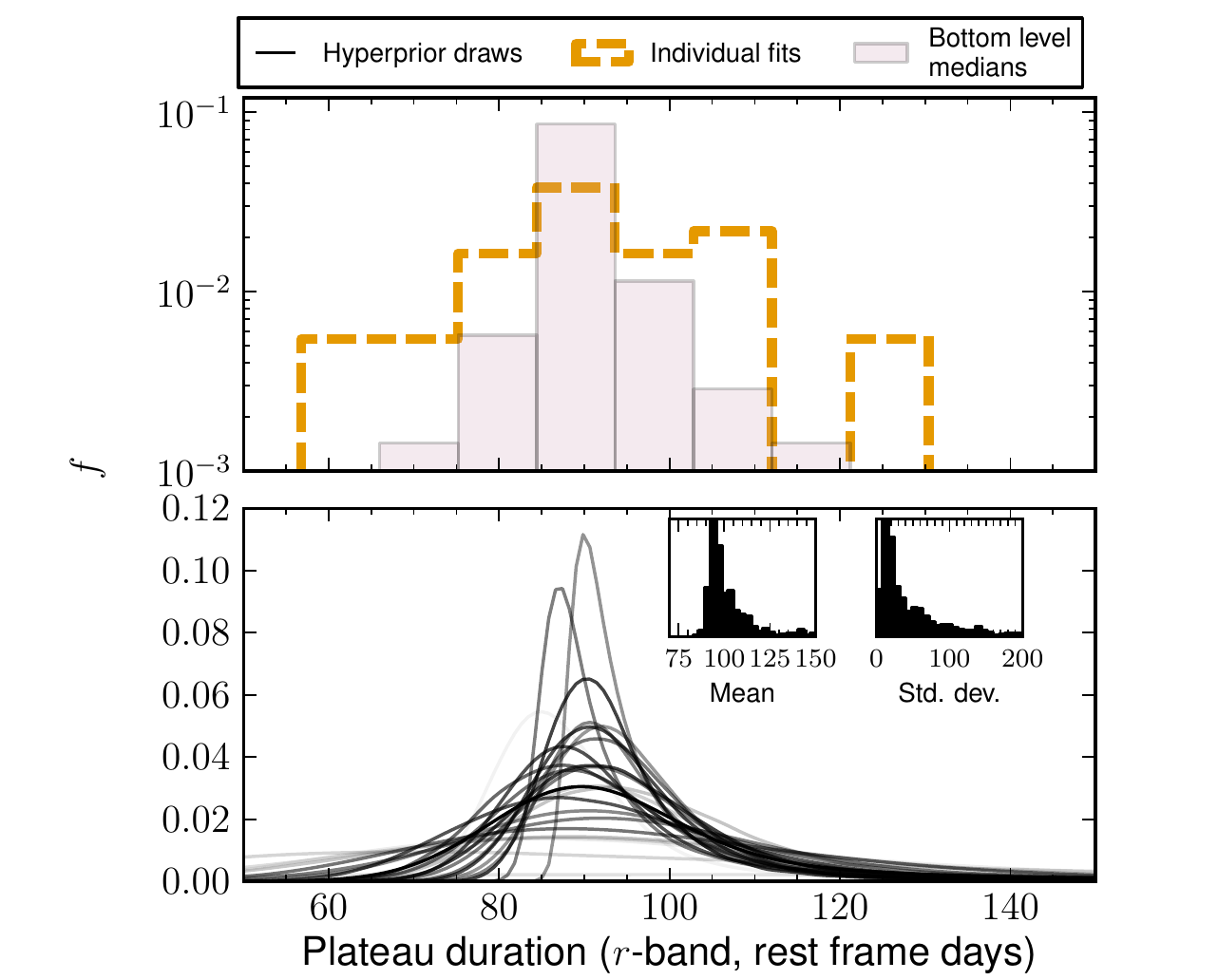}
\caption{\label{fig:tplatdist}Comparison of population level inferences on the SN~IIP plateau duration distributions (in $r$-band) between the hierarchical and individual model fits.  Top: The dashed line shows the histogram of plateau durations measured in \cite{Sanders14IIP} using the individual model (including only the objects with well constrained posteriors, as defined therein).  The shaded area shows a histogram of the bottom level plateau duration parameters under the hierarchical model (for all objects in the sample).  Bottom: The solid lines show draws of the plateau distribution hyperprior (a sum over the $t_p$ and $t_2$ hyperpriors) from the hierarchical model fit, with the opacity scaled to the probability of the draw under the posterior.  The inset plots show the marginalized posterior distributions for the mean and standard deviation of the summed hyperprior distributions (in rest frame days).}
\end{figure*}

To further the investigation of SN~IIP plateau durations from \cite{Sanders14IIP}, in Figure~\ref{fig:tplatdist} we compare the plateau duration inferences from that project to the duration distributions inferred from the hierarchical model.  The plateau duration distribution hyperprior in the hierarchical model is a sum of the $t_p$ and $t_2$ hyperpriors. The hyperpriors include both location (e.g. $t_{hP}$) and width (e.g. $\sigma~t_{hP}$) hyperparameters, so we visualize the posterior distribution of hyperpriors by showing multiple lognormal hyperpriors corresponding to random draws of the hyperparameters.  We focus on the $r$-band durations here, and so include both the $t_{hP}$ and $t_{hF}$ hyperparameters.  

The distribution of bottom-level plateau duration parameters for our hierarchical model agree well with the individual light curve fits from \cite{Sanders14IIP}.  Taking the median bottom level parameter from the MCMC chain, the distribution of values from the hierarchical model has a mean and standard deviation of $\tplatHIERdistmean\pm\tplatHIERdiststd$~days, compared to $\tplatINDdistmean\pm\tplatINDdiststd$~days for the individual model.  Note that the variance in the hierarchical model distribution is significantly lower than from the individual fits, because partial pooling between objects constrains the bottom level posteriors.

By directly modeling the underlying population of transients, the hierarchical modeling framework allows us to overcome potential biases in transient search methodology.  In particular, although long duration SNe~IIP are less likely to be observed with full temporal coverage in ground based transient searches (Section~\ref{sec:model}), we can estimate the fraction of unseen, long-duration transients in the population from the hierarchical model posterior.  This product of the posterior distribution of the hierarchical model is constrained by both the observed characteristics of objects in the sample, and characteristics allowed by pre-explosion and late time non-detections from the transient search at the same location.

Our hyperparameter posterior distributions suggest there is a \tplatAboveMaxPten\% probability that at least 10\% of the underlying population of SNe~IIP have $r$-band plateau durations longer than the bottom level parameter value for any individual object in the sample ($>\tplatHIERdistmax$~days).  The probability that at least 20\% of objects fall above this value is \tplatAboveMaxPtwen\%.  Among the sampled hyperprior distributions, the median of the population standard deviation is \tplatHIERdiststdmed~days.  The standard deviation distribution has a strong tail at larger values, shown in Figure~\ref{fig:tplatdist}.  These results emphasize and support the finding of \cite{Sanders14IIP}, that the plateau duration distribution of SNe~IIP has significant variance.

\section{DISCUSSION}
\label{sec:disc} 

The multi-level model developed and applied in this paper points to a methodological framework for the interpretation of SN light curves from the next generation of wide-field transient searches, such as the Large Synoptic Survey Telescope (LSST; \citealt{LSSTbook}).  In the coming era, the volume of available photometric data will increase dramatically, while the human and observational resources for follow-up of individual objects will not.  In this regime, to capitalize on the larger SN sample sizes afforded by these next generation searches, it will be critical to apply population level light curve modeling.  To do this, it is necessary to apply analytical methods that are robust to data sparseness and posterior non-identifiability for individual objects, and computational methods that are capable of generating inferences from large datasets.  The combination of hierarchical Bayesian methodology and Hamiltonian Monte Carlo methods explored in this work are natural methods for addressing both these concerns.

This work also suggests future paths for improvement of the Bayesian light curve modeling framework presented here.  First, to permit applications to purely photometric datasets (where the SN classification and redshift are not known or poorly constrained), the model must be generalized.  The redshift can trivially be added to the multi-level model as a vector of free parameters, but it will introduce significant posterior correlations and interactions that will need to be tested and may require the assertion of significant prior information to aid identifiability.  Additional model components are needed to allow for application to multiple SN classes simultaneously.  This could be accomplished through categorical mixture modeling with several different, physically-motivated SN light curve prescriptions for each SN type or by using a more generic, non-parametric, or continuously expanding light curve model to allow fitting of diverse SN types.  Both of these solutions would support classification inferences, by assessment of the categorical simplex parameter posteriors directly, or by a clustering analysis of the continuous model expansion parameters, respectively.  Applications to transient searches delving to significantly higher redshifts (and also the non-parametric modeling approach) may require full three-dimensional modeling of the SN spectral energy distribution evolution, rather than two-dimensional light curve modeling in each filter, in order to permit $K$-corrections at all distances.  Finally, incorporating host galaxy information will be critical to producing purely-photometric informative inferences across SN classes.  This should include modeling of the distribution of host galaxy global properties per SN class to uncover and take advantage of differences in progenitor properties (see e.g. \citealt{Kelly12,SandersIbc,galsnid,Lunnan13b,McCrum14}), as well as the line-of-sight host galaxy extinction and reddening law (e.g. \citealt{Mandel11}).  

Finally, we look to Riemannian manifold Hamiltonian Monte Carlo (RMHMC; \citealt{betancourt:2013}) techniques for permitting posterior characterization in the future, in the face of these additional modeling complexities.  Compared to traditional, Euclidian Hamiltonian Monte Carlo (EHMC; as applied for this work), RHMC samplers efficiently explore highly correlated and high-dimensional posterior functions by automatically adapting the Hamiltonian integration step size to a value optimal for local conditions \citep{Girolami11,Betancourt13}.  This capability would permit unbiased sampling even amidst models with joint posterior distributions with higher curvature than the multi-level model examined here, such as a model including interactions between filter-level parameters or interactions between light curve parameter groups (e.g. $t$-$r$ interactions).  Our attempts to fit such a model with EHMC have not achieved convergence within reasonable integration times, with the high posterior curvature preventing the hyperparameters from moving at a sufficient rate to produce convergent chains.  The addition of RHMC sampling capabilities to \textit{Stan} in the near future \citep{betancourt:2013} will make these techniques accessible to astronomers in the LSST era.

\section{CONCLUSIONS}
\label{sec:conc} 

We have explored the use of Bayesian hierarchical modeling and Hamiltonian Monte Carlo (HMC) to enable population-level inference on multi-band transient light curves from comprehensive analysis of optical photometry from wide field transient searches.  The primary conclusions of this work are:

\begin{itemize}
\item While computational limits still challenge the implementation of hierarchical models, due to the high curvature in their joint posterior distributions, sufficient convergence is achieved in the bottom level model parameters (Section~\ref{sec:res:converge}) to enable their immediate application for transient light curve studies.
\item Comparisons between light curve posterior predictive distributions from our hierarchical model fit to the individual light curve fits of \cite{Sanders14IIP} show strong agreement for well identified parameters, and show an advantage for hierarchical models among poorly identified parameters (Section~\ref{sec:res:PPC}).  In particular, partial pooling of parameter information between transients supports improved regularization of light curve shapes, and supports model selection between partially degenerate light curve parameter scenarios.
\item By directly modeling the underlying transient population, hierarchical models permit inference on the occurrence of properties not observed within the dataset (Section~\ref{sec:res:pop}).  This feature is of particular value in overcoming observational biases induced by ground based transient searches, such as the under-representation of long duration transients like some SNe~IIP.
\end{itemize}

We have concluded with a discussion of future directions for this modeling (Section~\ref{sec:disc}), including applications to upcoming wide field transient searches, extensions to the hierarchical model structure developed here, and expanded capabilities to be enabled by the advent of Riemannian manifold Hamiltonian Monte Carlo.

\acknowledgements
\label{sec:ackn}

We thank K. Mandel for sage guidance and many helpful conversations; M. Brubaker, B. Carpenter, A. Gelman, and the \textit{Stan} team for their excellent modeling language and HMC sampler and for thoughtful feedback on our model design; and the entire PS1 collaboration for their monumental efforts towards the collection of the SN~IIP light curve dataset.

The Pan-STARRS1 Surveys (PS1) have been made possible through contributions of the Institute for Astronomy, the University of Hawaii, the Pan-STARRS Project Office, the Max-Planck Society and its participating institutes, the Max Planck Institute for Astronomy, Heidelberg and the Max Planck Institute for Extraterrestrial Physics, Garching, The Johns Hopkins University, Durham University, the University of Edinburgh, Queen's University Belfast, the Harvard-Smithsonian Center for Astrophysics, the Las Cumbres Observatory Global Telescope Network Incorporated, the National Central University of Taiwan, the Space Telescope Science Institute, the National Aeronautics and Space Administration under Grant No. NNX08AR22G issued through the Planetary Science Division of the NASA Science Mission Directorate, the National Science Foundation under Grant No. AST-1238877, the University of Maryland, and Eotvos Lorand University (ELTE).

Support for this work was provided by the David and Lucile Packard Foundation Fellowship for Science and Engineering awarded to A.M.S.  M.B. is supported under EPSRC grant EP/J016934/1.  Computations presented in this paper were performed using the Odyssey supercomputing cluster supported by the FAS Science Division Research Computing Group at Harvard University.

{\it Facilities:} \facility{PS1}

\bibliographystyle{fapj}

\clearpage

\begin{appendix}
\section{Appendix A: Hierarchical Light Curve Stan Model}
\label{ap:stan}

Below we reproduce the full hierarchical model for the 5 component piecewise SN~IIP light curves in the  \textit{Stan} modeling language, as described in Section~\ref{sec:model}.  The Stan model specification format is documented in the Stan Modeling Language Users Guide and Reference Manual \citep{stan-manual:2014}.  

The model takes the following data as input: \texttt{N\_obs}, the total number of photometric data points; \texttt{N\_filt}, the number of photometric filters; \texttt{t}, a vector of MJD dates of the photometric observations; \texttt{fL}, a vector of luminosities corresponding to the photometric observations (with units as described in \citealt{Sanders14IIP}); \texttt{dfL} a corresponding vector of luminosity uncertainties; \texttt{z} the redshift; \texttt{t0\_mean} an initial estimate of the explosion date (for initialization and for centering the explosion date prior distribution); \texttt{J} a vector of integers specifying the filter ID of each photometric observation; \texttt{Kcor\_N}, a matrix of pre-computed $K$-corrections for each filter, in magnitudes with spacing of 1~day; \texttt{fluxscale} the zero-point of the luminosity unit system ($\texttt{fluxscale} = 10^7$ in the system we have employed); and \texttt{duringseason}, a boolean value specifying whether the object exploded within or between observing seasons, for selection of the explosion date prior distribution parameters.  The calculation of the model light curve flux and application of the $K$-correction values is performed in the \texttt{transformed parameters} section, and the prior and likelihood calculations are performed in the \texttt{model} section.  Certain vector-valued prior distribution parameters are specified in the \texttt{transformed data} section for convenience.  We note that the higher level parameters for the five different light curve rates and four different phase durations are grouped together in vectors (e.g. $r_{hP}$ and $t_{hP}$ for the top level, and $r_{hF}$ and $t_{hF}$ for the filter level, respectively) for convenience.

The \textit{Stan} model is then compiled and run \citep{stan-manual:2014} to yield MCMC samples from the posterior distribution of light curve parameters.  We configured the \textit{No-U-Turn Sampler} to use fixed $0$ initialization of the parameter values, an adaptation phase of 30~steps, a maximum treedepth of 16, and otherwise employed the default sampler parameters.  We have used CmdStan version 2.2.0\footnote{\url{https://github.com/stan-dev/stan/releases/tag/v2.2.0}}.

{\footnotesize
\begin{verbatim}
data {
  int<lower=0> N_obs;                             
  int<lower=0> N_SN;                              
  int<lower=0> N_filt;                            
  vector[N_obs] t;                                
  vector[N_obs] fL;                               
  vector[N_obs] dfL;                              
  vector[N_SN] z;                                 
  vector[N_SN] t0_mean;                           
  int<lower=1,upper=N_filt> J[N_obs];             
  int<lower=1,upper=N_SN> SNid[N_obs];            
  int<lower=0> Kcor_N;                            
  real Kcor[N_SN, N_filt,Kcor_N];                 
  real<lower=0> fluxscale;                        
  vector<lower=0,upper=1>[N_SN] duringseason;     
}
transformed data {
  vector[N_filt] prior_t_hF[4];                           
  vector[N_filt] prior_t_hF_s[4];
  vector[N_filt] prior_r_hF[5];                           
  vector[N_filt] prior_r_hF_s[5];
  for (i in 1:N_filt) {
    prior_t_hF[1,i] <- 0;
    prior_t_hF_s[1,i] <- 0.1;
  }
  prior_t_hF[2,1] <- -1;
  prior_t_hF[2,2] <- -0.5;
  prior_t_hF[2,3] <- 0;
  prior_t_hF[2,4] <- 0.5;
  prior_t_hF[2,5] <- 1;
  for (i in 1:N_filt) {prior_t_hF_s[2,i] <- 0.1;}
  for (i in 1:N_filt) {
    prior_t_hF[3,i] <- 0;
    prior_t_hF_s[3,i] <- 0.1;
  }
  for (i in 1:N_filt) {
    prior_t_hF[4,i] <- 0;
    prior_t_hF_s[4,i] <- 0.1;
  }
  for (i in 1:N_filt) {
    prior_r_hF[1,i] <- 0;
    prior_r_hF_s[1,i] <- 0.1;
  }
  prior_r_hF[2,1] <- 2;
  prior_r_hF[2,2] <- 1;
  prior_r_hF[2,3] <- 0;
  prior_r_hF[2,4] <- -0.5;
  prior_r_hF[2,5] <- -1;
  for (i in 1:N_filt) {prior_r_hF_s[2,i] <- 0.1;}
  prior_r_hF[3,1] <- 1;
  prior_r_hF[3,2] <- 0.3;
  prior_r_hF[3,3] <- 0;
  prior_r_hF[3,4] <- -1;
  prior_r_hF[3,5] <- -1;
  for (i in 1:N_filt) {prior_r_hF_s[3,i] <- 0.1;}
  for (i in 1:N_filt) {
    prior_r_hF[4,i] <- 0;
    prior_r_hF_s[4,i] <- 0.1;
  }
  for (i in 1:N_filt) {
    prior_r_hF[5,i] <- 0;
    prior_r_hF_s[5,i] <- 0.1;
  }
}
parameters {  
  vector[4] t_hP;                                 
  vector<lower=0>[4] sig_t_hP;                    
  vector[N_filt] t_hF[4];                         
  vector<lower=0>[N_filt] sig_t_hF[4];            
  vector[N_SN * N_filt] t_hSNF[4];                
  vector<lower=0>[N_SN * N_filt] sig_t_hSNF[4];   
  vector[5] r_hP;                                 
  vector<lower=0>[5] sig_r_hP;                    
  vector[N_filt] r_hF[5];                         
  vector<lower=0>[5] sig_r_hF[5];                 
  vector[N_SN * N_filt] r_hSNF[5];                
  vector<lower=0>[N_SN * N_filt] sig_r_hSNF[5];   
  real M_h;                                       
  real<lower=0> sig_M_h;                          
  vector[N_filt] M_hF;                            
  vector<lower=0>[N_filt] sig_M_hF;               
  vector[N_SN * N_filt] M_hSNF;                   
  vector<lower=0>[N_SN * N_filt] sig_M_hSNF;      
  real Y_h;                                      
  real<lower=0> sig_Y_h;                         
  vector[N_SN * N_filt] Y_hSNF;                    
  vector<lower=0>[N_SN * N_filt] sig_Y_hSNF;       
  real t0s_h;                                   
  real<lower=0> sig_t0s_h;                      
  vector[N_SN] t0s_hSN;                         
  vector<lower=0>[N_SN] sig_t0s_hSN;            
  real t0l_h;                                   
  real<lower=0> sig_t0l_h;                      
  vector[N_SN] t0l_hSN;                         
  vector<lower=0>[N_SN] sig_t0l_hSN;            
  real<lower=0> V_h;                            
  vector<lower=0>[N_filt] V_hF;                 
  vector<lower=0>[N_SN * N_filt] V_hSNF;          
}
transformed parameters {
    vector[N_obs] mm;                         
    vector[N_obs] dm;                         
    vector<upper=0>[N_SN] pt0;
    matrix<lower=0>[N_SN, N_filt] t1;
    matrix<lower=0>[N_SN, N_filt] t2;
    matrix<lower=0>[N_SN, N_filt] td;
    matrix<lower=0>[N_SN, N_filt] tp;
    matrix[N_SN, N_filt] lalpha; 
    matrix[N_SN, N_filt] lbeta1;
    matrix[N_SN, N_filt] lbeta2;
    matrix[N_SN, N_filt] lbetadN;
    matrix[N_SN, N_filt] lbetadC; 
    matrix[N_SN, N_filt] Mp; 
    matrix[N_SN, N_filt] Yb;
    matrix<lower=0>[N_SN, N_filt] V;
    matrix<lower=0>[N_SN, N_filt] M1;
    matrix<lower=0>[N_SN, N_filt] M2;
    matrix<lower=0>[N_SN, N_filt] Md;
    for (l in 1:N_SN) {
      if (duringseason[l] == 1) {
        pt0[l] <- -exp( t0s_h + sig_t0s_h * ( t0s_hSN[l] .* sig_t0s_hSN[l] )); 
      } else {
        pt0[l] <- -exp( t0l_h + sig_t0l_h * ( t0l_hSN[l] .* sig_t0l_hSN[l] ));
      }
    }
    for (i in 1:N_filt) { 
	
        for (j in 1:N_SN) { 
            t1[j,i] <- exp( log(1) + t_hP[1] + sig_t_hP[1] * (
                       t_hF[1,i] * sig_t_hF[1,i] 
                     + sig_t_hSNF[1,(i-1)*N_SN+j] * t_hSNF[1,(i-1)*N_SN+j]
                       ));
	    
            tp[j,i] <- exp( log(10) + t_hP[2] + sig_t_hP[2] * (
                       t_hF[2,i] * sig_t_hF[2,i] 
                     + sig_t_hSNF[2,(i-1)*N_SN+j] * t_hSNF[2,(i-1)*N_SN+j]
                       ));
            t2[j,i] <- exp( log(100) + t_hP[3] + sig_t_hP[3] * (
                       t_hF[3,i] * sig_t_hF[3,i] 
                     + sig_t_hSNF[3,(i-1)*N_SN+j] * t_hSNF[3,(i-1)*N_SN+j]
                       ));
            td[j,i] <- exp( log(10) + t_hP[4] + sig_t_hP[4] * (
                       t_hF[4,i] * sig_t_hF[4,i] 
                     + sig_t_hSNF[4,(i-1)*N_SN+j] * t_hSNF[4,(i-1)*N_SN+j]
                       ));
            lalpha[j,i] <- -1 + ( r_hP[1] + sig_r_hP[1] * (
                           r_hF[1,i] * sig_r_hF[1,i] 
                         + sig_r_hSNF[1,(i-1)*N_SN+j] * r_hSNF[1,(i-1)*N_SN+j]
                           ));
            lbeta1[j,i] <- -4 + ( r_hP[2] + sig_r_hP[2] * (
                           r_hF[2,i] * sig_r_hF[2,i] 
                         + sig_r_hSNF[2,(i-1)*N_SN+j] * r_hSNF[2,(i-1)*N_SN+j]
                           ));
            lbeta2[j,i] <- -4 + ( r_hP[3] + sig_r_hP[3] * (
                           r_hF[3,i] * sig_r_hF[3,i] 
                         + sig_r_hSNF[3,(i-1)*N_SN+j] * r_hSNF[3,(i-1)*N_SN+j]
                           ));
            lbetadN[j,i] <- -3 + ( r_hP[4] + sig_r_hP[4] * (
                            r_hF[4,i] * sig_r_hF[4,i] 
                          + sig_r_hSNF[4,(i-1)*N_SN+j] * r_hSNF[4,(i-1)*N_SN+j]
                            ));
            lbetadC[j,i] <- -5 + ( r_hP[5] + sig_r_hP[5] * (
                            r_hF[5,i] * sig_r_hF[5,i] 
                          + sig_r_hSNF[5,(i-1)*N_SN+j] * r_hSNF[5,(i-1)*N_SN+j]
                            ));
            Mp[j,i] <- exp(M_h + sig_M_h * (
                           M_hF[i] * sig_M_hF[i] 
                         + sig_M_hSNF[(i-1)*N_SN+j] * M_hSNF[(i-1)*N_SN+j]
                           ));
            Yb[j,i] <- Y_h + sig_Y_h * (Y_hSNF[(i-1)*N_SN+j] .* sig_Y_hSNF[(i-1)*N_SN+j]);
            V[j,i] <- V_h * V_hF[i] * V_hSNF[(i-1)*N_SN+j]; 
        }
    }
    M1 <- Mp ./ exp( exp(lbeta1) .* tp );
    M2 <- Mp .* exp( -exp(lbeta2) .* t2 );
    Md <- M2 .* exp( -exp(lbetadN) .* td );
    for (n in 1:N_obs) {      
        real N_SNc;                                      
        int Kc_up;                                       
        int Kc_down;                                     
        real t_exp;                                      
        int j;                                           
        int k;                                           
        real mm_1;
        real mm_2;
        real mm_3;
        real mm_4;
        real mm_5;
        real mm_6;
        j <- J[n];
        k <- SNid[n];
        t_exp <- ( t[n] - (t0_mean[k] + pt0[k]) ) / (1 + z[k]);
        if (t_exp<0) {                                                                      
            mm_1 <- Yb[k,j];
        } else {
            mm_1 <- 0;
        }
	if ((t_exp>=0) && (t_exp < t1[k,j])) {                                                                      
            mm_2 <- Yb[k,j] + M1[k,j] * pow(t_exp / t1[k,j] , exp(lalpha[k,j]));
        } else {
            mm_2 <- 0;
        }
        if ((t_exp >= t1[k,j]) && (t_exp < t1[k,j] + tp[k,j])) {                                                                      
            mm_3 <- Yb[k,j] + M1[k,j] * exp(exp(lbeta1[k,j]) * (t_exp - t1[k,j]));
        } else {
            mm_3 <- 0;
        }
        if ((t_exp >= t1[k,j] + tp[k,j]) && (t_exp < t1[k,j] + tp[k,j] + t2[k,j])) {                                                                      
            mm_4 <- Yb[k,j] + Mp[k,j] * exp(-exp(lbeta2[k,j]) * (t_exp - t1[k,j] - tp[k,j]));
        } else {
            mm_4 <- 0;
        }
        if ((t_exp >= t1[k,j] + tp[k,j] + t2[k,j]) && (t_exp < t1[k,j] + tp[k,j] + t2[k,j] + td[k,j])) {                                                                      
            mm_5 <- Yb[k,j] + M2[k,j] * exp(-exp(lbetadN[k,j]) * (t_exp - t1[k,j] - tp[k,j] - t2[k,j]));
        } else {
            mm_5 <- 0;
        }
        if (t_exp >= t1[k,j] + tp[k,j] + t2[k,j] + td[k,j]) {                                                                      
            mm_6 <- Yb[k,j] + Md[k,j] * exp(-exp(lbetadC[k,j]) * (t_exp - t1[k,j] - tp[k,j] - t2[k,j] - td[k,j]));
        } else {
            mm_6 <- 0;
        }
        dm[n] <- sqrt(pow(dfL[n],2) + pow(V[k,j],2));
        if (t_exp<0) {
            N_SNc <- 0;
        } else if  (t_exp<Kcor_N-2){                    
            Kc_down <- 0;
            while ((Kc_down+1) < t_exp) {               
		Kc_down <- Kc_down + 1; 
            }
            Kc_up <- Kc_down+1;
            N_SNc <- Kcor[k,j,Kc_down+1] + (t_exp - floor(t_exp)) * (Kcor[k,j,Kc_up+1]-Kcor[k,j,Kc_down+1]);
        } else {                                        
            N_SNc <- Kcor[k,j,Kcor_N];
        }
        mm[n] <- (mm_1+mm_2+mm_3+mm_4+mm_5+mm_6) / (pow(10, N_SNc/(-2.5)));
    }
}
model {
    t0s_h ~ normal(0, 0.5);                             
    sig_t0s_h ~ cauchy(0, 0.1);
    t0l_h ~ normal(log(100), 1);                       
    sig_t0l_h ~ cauchy(0, 0.1);
    V_h ~ cauchy(0, 0.001);                               
    Y_h ~ normal(0, 0.1);                               
    sig_Y_h ~ cauchy(0, 0.01);
    M_h ~ normal(0, 1);                                
    sig_M_h ~ cauchy(0, 0.1);
    t_hP ~ normal(0,0.1);                             
    sig_t_hP ~ cauchy(0, 0.1);
    for (i in 1:4) {                                  
      t_hF[i] ~ normal(prior_t_hF[i], prior_t_hF_s[i]);
      sig_t_hF[i] ~ cauchy(0, 0.1);
      t_hSNF[i] ~ normal(0,1);
      sig_t_hSNF[i] ~ cauchy(0, 0.1);
    }
    r_hP ~ normal(0,1);                              
    sig_r_hP ~ cauchy(0, 0.1);
    for (i in 1:5) {
      r_hF[i] ~ normal(prior_r_hF[i], prior_r_hF_s[i]);
      sig_r_hF[i] ~ cauchy(0, 0.1);
      r_hSNF[i] ~ normal(0,1);
      sig_r_hSNF[i] ~ cauchy(0, 0.1);
    }
    M_hF ~ normal(0,1);
    sig_M_hF ~ cauchy(0, 0.1);
    M_hSNF ~ normal(0,1);
    sig_M_hSNF ~ cauchy(0, 0.1);
    Y_hSNF ~ normal(0,1);                       
    sig_Y_hSNF ~ cauchy(0, 0.1);
    V_hF ~ cauchy(0, 0.1);
    V_hSNF ~ cauchy(0, 0.1);
    t0s_hSN ~ normal(0,1);                           
    sig_t0s_hSN ~ cauchy(0, 0.1);
    t0l_hSN ~ normal(0,1);
    sig_t0l_hSN ~ cauchy(0, 0.1);
    fL ~ normal(mm,dm);
}
\end{verbatim}
}

\end{appendix}

\end{document}